\documentclass[journal]{IEEEtran}

\usepackage{latexsym}
\usepackage{graphicx}
\usepackage{amsfonts,amssymb,amsmath}
\usepackage{ctable} 
\usepackage{mathtools, cuted}
\usepackage{hyperref}
\usepackage[ruled,linesnumbered,noend]{algorithm2e}
\usepackage[T1]{fontenc}
\usepackage{cite}
\usepackage{subcaption}
\usepackage{comment}
\usepackage{amsthm}
\usepackage{overpic}
\usepackage{steinmetz}
\usepackage{array}
\usepackage{url}
\usepackage{color}
\usepackage{algorithmicx}
\usepackage{algcompatible}
\usepackage{xcolor}
\usepackage{soul}
\usepackage{flushend}

\theoremstyle{plain}

\newcommand{\argmin}[1]{{\underset{{#1}}{\mathrm{arg\,min}}}}
\newcommand{\vect}[1]{\mathbf{#1}}

\newcommand{\minimize}[1]{{\underset{{#1}}{\mathrm{minimize}}}}

\newcommand{\bl}[1]{\boldsymbol{#1}}

\def\diag{\mathrm{diag}}

\def\Ttran{\mbox{\tiny $\mathrm{T}$}}
\def\CN{\mathcal{N}_{\mathbb{C}}} 
\def\T{\mathrm{T}}
\def\H{\mathrm{H}}

\def\mod{\mathrm{mod}}

\begin{document}

\title{MSE Minimization in RIS-Aided MU-MIMO with Discrete Phase Shifts and Fronthaul Quantization}

\author{Parisa Ramezani, Yasaman Khorsandmanesh, and Emil Bj\"{o}rnson\\
\IEEEauthorblockA{\textit{ Department of Computer Science, KTH Royal Institute of Technology, Stockholm, Sweden} \\  Email: \{parram, yasamank, emilbjo\}@kth.se}}

\maketitle

\begin{abstract}
    In this paper, we consider a downlink multi-user multiple-input multiple-output  (MU-MIMO)  communication assisted by a reconfigurable intelligent surface (RIS) and study the precoding and RIS configuration design under practical system constraints. These constraints include the limited-capacity fronthaul at the transmitter side and the finite resolution of RIS elements. We investigate the sum mean squared error (MSE) minimization problem and propose an algorithm based on the block coordinate descent method to optimize the precoding, RIS configuration, and receiver gains. We compute the precoding vectors and RIS configuration using the Schnorr-Euchner sphere decoding (SESD) method which delivers the optimal MSE-minimizing solution. We numerically evaluate the performance of the proposed SESD-based methods and corroborate their effectiveness in improving the system performance.
\end{abstract}

\section{Introduction}

In recent years, reconfigurable intelligent surfaces (RISs) have emerged as a new design  in wireless communication systems. These tunable surfaces are composed of a multitude of electrically small reflective components, capable of altering the electromagnetic properties of the incident signals to control the propagation environment. 
This revolutionary technology offers promising solutions to address the escalating demand for higher data rates and improved energy efficiency in next-generation wireless networks \cite{Pan2020multicell}. Thus, exploring RIS's theoretical foundations and practical implementations is critical to leveraging its full potential. One practical limitation of RIS is its finite bit resolution, which means that each element can only select a phase shift from a discrete set.
A further practical system constraint is related to the limited fronthaul capacity. As the bandwidth and antenna numbers increase, the load on the fronthaul between the advanced antenna system (AAS) on the base station (BS) site and the cloud-based baseband unit (BBU) becomes a limiting factor. 
Therefore, the results of computations performed at the BBU must be quantized before being sent to the AAS via the fronthaul. 

The two above-mentioned constraints have been investigated in some recent works \cite{Wu2019b,Di2020,Ramezani2023Novel,khorsandmanesh2023optimized,khorsandmanesh2023fronthaul}. Specifically, considering discrete phase shifts at the RIS, the papers \cite{Wu2019b,Di2020} obtained sub-optimal discrete phase shifts at the RIS by quantizing an optimal continuous phase shift vector, while the authors in \cite{Ramezani2023Novel} proposed a novel phase shift design for finding the optimal discrete phase shifts at the RIS in an uplink multi-user multiple-input multiple-output (MU-MIMO) setup. With regard to the fronthaul capacity limitation and the need for quantizing the precoding vectors, papers \cite{khorsandmanesh2023optimized} and \cite{khorsandmanesh2023fronthaul} introduced innovative precoding strategies targeting the minimization of the sum mean squared error (MSE) and the maximization of the sum rate within an MU-MIMO framework, respectively. Although the RIS phase shift and fronthaul limitations have been separately investigated in these recent works, there has been little attention to their joint effect in RIS-aided MIMO systems.

This paper considers an RIS-aided downlink MU-MIMO system. 
We consider a challenging MSE minimization problem, aiming to determine the optimal discrete precoding vectors, discrete RIS phase shifts, and receiver gains. We present a new block coordinate descent algorithm that alternately optimizes these variables. Although the block coordinate descent method is prevalent in multi-variable optimization, it is often overlooked that the convergence to a stationary point is only guaranteed when each sub-problem is solved optimally \cite{Tseng2001}. In this paper, we solve all the three sub-problems optimally. 
Specifically, we address the optimization of the discrete precoding vectors and RIS phase shifts by formulating them as mixed integer least squares problems and introduce a novel algorithm based on the 
Schnorr-Euchner sphere decoding (SESD) technique to obtain the optimal discretized precoding and  RIS phase shifts. The optimal closed-form solution for the receiver gains is also presented.  
 Through numerical simulations, we demonstrate the importance of obtaining the optimal discrete precoding and RIS configuration.
 
\section{System Model}
\label{sec:SysMod}
We consider a downlink RIS-assisted MU-MIMO communication system, as illustrated in Fig.~\ref{fig:system}, where a BS equipped with an AAS with $M$ antennas intends to transmit information to $K$ single-antenna user equipments (UEs). The direct links between the BS and the UEs are blocked. However, an RIS with $N$ reflecting elements is deployed to assist the communication. The AAS is connected to the BBU through a limited-capacity digital fronthaul. The signal intended for UE\,$k$ (henceforth denoted by $\mathrm{U}_k$) consists of a precoding vector and a data symbol. Since the data symbols are bit sequences picked from a channel coding codebook, the BBU can send them to the AAS via the fronthaul link without quantization errors. On the other hand, the precoding vectors normally contain arbitrary complex-values entries and must be quantized to finite resolution before being sent over the fronthaul. The quantized precoding vectors are multiplied by the corresponding data symbols at the AAS and the product is transmitted to the UEs via RIS. The signal received at $\mathrm{U}_k$ is given by 
\begin{figure}[t!]
  \centering
   \begin{overpic}[scale = 0.28]{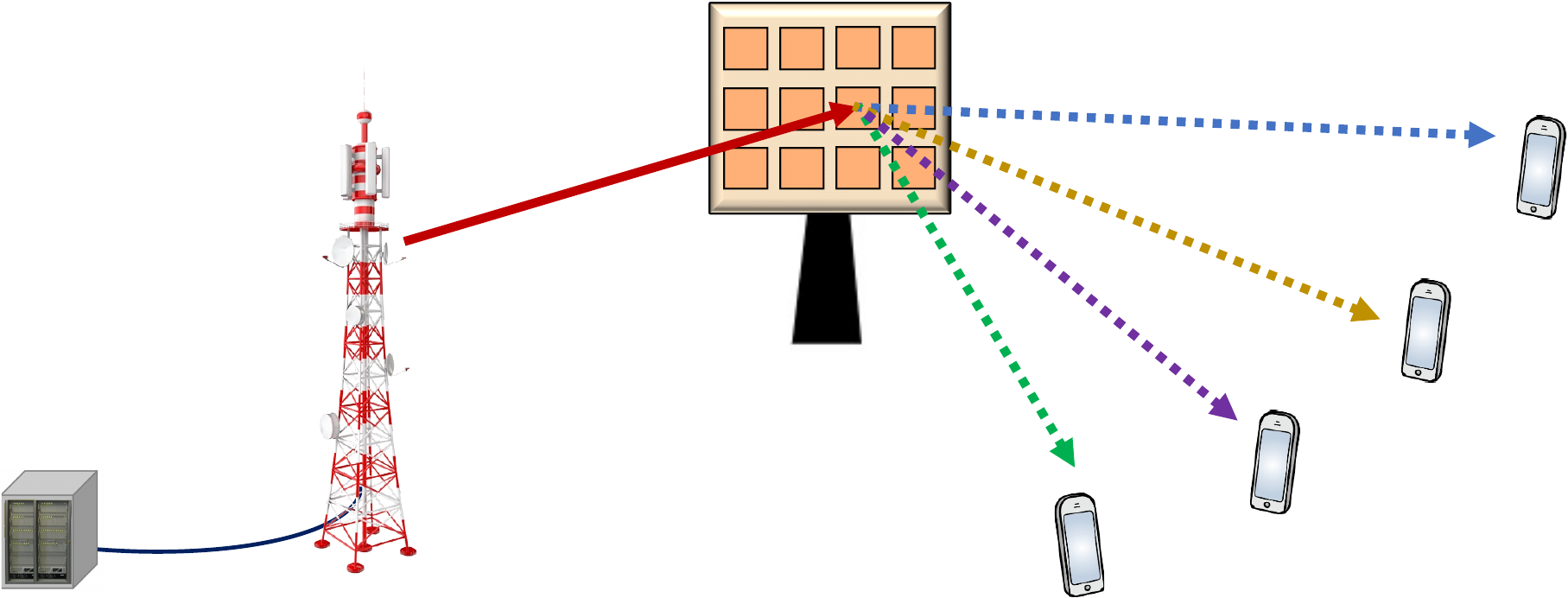}
  \put(0,9){\footnotesize BBU}%
  \put(20,32){\footnotesize AAS}%
  \put(50,40){\footnotesize RIS}%
  \put(80,1){\footnotesize UE\,$k$}%
  \put(35,27){\footnotesize $\vect{H}$}
  \put(70,20){\footnotesize $\vect{g}_k$}
   \end{overpic}
\caption{An RIS-assisted MU-MIMO downlink communication.}
\label{fig:system}
\end{figure}
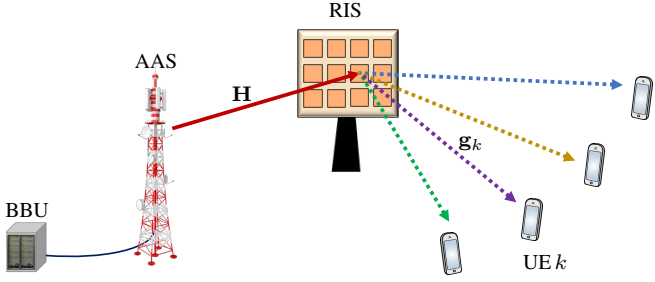
\begin{equation}
 y_k = \vect{g}_k^\T \bl{\Theta}\vect{H} \left(\sum_{i=1}^K \vect{w}_i s_i \right) + n_k,   
\end{equation}where $\vect{g}_k \in \mathbb{C}^N$ is the channel between the RIS and $\mathrm{U}_k$, $\vect{H} \in \mathbb{C}^{N \times M}$ is the channel between the BS and the RIS, and $\bl{\Theta}$ is an $N \times N$ diagonal matrix having the RIS reflection coefficients on its main diagonal. Specifically, $\bl{\Theta} = \diag([\theta_1,\ldots,\theta_N]^{\Ttran} )$ with $\theta_n = e^{j\vartheta_n}$ and $\vartheta_n$ representing the phase shift applied to the incident signal by the $n$th RIS element. Furthermore, $s_i$ is the unit-power data symbol of $\mathrm{U}_i$ and $n_k \sim \CN (0,N_0)$ indicates the independent additive complex Gaussian noise with power $N_0$. Finally, $\vect{w}_i \in \mathcal{P}^{M}$ is the precoding vector applied to $\mathrm{U}_i$'s data symbol where $\mathcal{P}$ is the fronthaul quantization alphabet. The transmitted signals must satisfy the total power constraint
\begin{equation}
    \label{eq:sum_power_constraint}
   \mathbb{E}\left\{\|\vect{w}_k s_k\|^2\right\} = \sum_{k=1}^K \|\vect{w}_k\|^2 \leq P,
\end{equation}where $P$ is the maximum average transmit power of the downlink signals and the equality is due to the fact that the transmitted symbols are independent and have unit power. The fronthaul quantization alphabet is defined as
\begin{equation}
\label{eq:quantization_alphabet}
    \mathcal{P} = \{l_R + jl_I: l_R,l_I \in \mathcal{L}\}.
\end{equation}
We assume that the same quantization alphabet is used for both real and imaginary components. In \eqref{eq:quantization_alphabet}, $\mathcal{L} = \left \{l_0,\ldots,l_{L-1} \right\}$ is the set of real-valued quantization labels. 

The limited-capacity fronthaul is modeled as a symmetric uniform quantizer.
The complex input $x = x_{\mathrm{R}} + jx_{\mathrm{I}}$ is mapped to $\mathcal{Q}(x) = l_p + jl_q$ where $l_p$ and $l_q$ are respectively the nearest values to $x_{\mathrm{R}}$ and $x_{\mathrm{I}} $ in set $\mathcal{L}$. We select the entries of $\mathcal{L}$ to minimize the distortion under the maximum entropy assumption \cite{Hui2001,khorsandmanesh2023optimized,khorsandmanesh2023fronthaul}.

Moreover, due to hardware limitations, the RIS elements can only pick phase shifts from a finite set of discrete values. Denoting by $b$ the number of quantization bits at the RIS, the set of discrete reflection coefficients is 
\begin{equation}
\label{eq:discrete_set}
    \mathcal{F} = \left\{e^{j\frac{m\pi}{2^{b-1}}}: m =0,1,\ldots,2^b - 1 \right\},
\end{equation}
$\mathrm{U}_k$ estimates its intended data symbol as $\hat{s}_k = \beta_k y_k$, where $\beta_k$ is the receiver gain. Setting $\bl{\theta} = [\theta_1,\ldots,\theta_N]^{\Ttran}$, the MSE between the actual and estimated data symbols is given by 
\begin{align}
\label{eq:MSE}
 e_k &= \mathbb{E} \left \{ |s_k - \hat{s}_k|^2 \right \} = |\beta_k|^2 \left ( \sum_{i=1}^K \left| \bl{\theta}^\T \diag(\vect{g}_k) \vect{H}\vect{w}_i\right|^2 + N_0 \right) \nonumber \\ 
 &- 2\Re \left ( \beta_k \bl{\theta}^\T \diag(\vect{g}_k) \vect{H} \vect{w}_k \right ) + 1.
\end{align}  
In this work, we focus on the design of precoding vectors, RIS configuration, and receiver gains to minimize the sum MSE of the UEs under the transmit power constraint in \eqref{eq:sum_power_constraint} and by taking into account the fronthaul quantization and finite resolution of RIS phase shifts. Our future work will feature a sum rate maximization problem where the equivalence between sum rate maximization and weighted sum MSE minimization will be utilized to maximize the sum rate. The sum MSE minimization problem is formulated as  
\begin{equation}
    \begin{aligned}
    \label{eq:main_problem}
        \minimize{ \bl{\beta} \in \mathbb{C}^K, \vect{W} \in \mathcal{P}^{M \times K}, \bl{\theta} \in \mathcal{F}^N}\,\, &\mathbb{E} \left \{ \| \vect{s} - \diag(\bl{\beta}) \vect{y}\|^2\right \} \\ \mathrm{subject~ to}\quad \quad \quad & \| \vect{W}\|_{\mathrm{F}} ^ 2 \leq P,    
    \end{aligned}
\end{equation}where $\bl{\beta} = [\beta_1,\ldots,\beta_K]^\T$, $\vect{W} = [\vect{w}_1,\ldots,\vect{w}_K]$, $\vect{s} = [s_1,\ldots,s_K]^\T$, $\vect{y} = [y_1,\ldots,y_K]^\T$, and $\|\cdot\|_{\mathrm{F}}$ denotes the Frobenius norm. 

\section{Proposed Solution}
Problem \eqref{eq:main_problem} is non-convex because the optimization variables are coupled in the objective function and several variables are discrete. To solve this problem, we employ the block coordinate descent method and optimize the precoding vectors, receiver gains, and RIS configuration alternately until convergence to a stationary point is achieved. 

   \subsection{Optimizing Precoding Vectors}
  Expanding the objective function in \eqref{eq:main_problem}, the precoding optimization problem can be re-written as 
  \begin{equation}
  \begin{aligned}
\label{eq:optimize_precoding}
     \minimize{\vect{W} \in \mathcal{P}^{M \times K}}\,\, &\sum_{k = 1}^K \vect{w}_k^\H  \vect{D}^\H \vect{D} \vect{w}_k - \vect{d}_k^\T \vect{w}_k - \vect{w}_k^\H \vect{d}_k^* \\
     \mathrm{subject~to}\,\,&\sum_{k=1}^K \vect{w}_k^\H \vect{w}_k \leq P,
     \end{aligned}
  \end{equation}where $\vect{D} = \left(\diag(\bl{\beta}) \otimes \bl{\theta}^\T  \right) \vect{G} \vect{H} \in \mathbb{C}^{K \times M} $,  $\vect{G} = \left[\diag(\vect{g}_1),\ldots,\diag(\vect{g}_K) \right]^\T \in \mathbb{C}^{KN \times N}$,
  and $\vect{d}_k^\T$ represents the $k$th row of $\vect{D}$. Furthermore, $(\cdot)^*$ indicates conjugation.
We use the method of Lagrange multipliers to solve Problem \eqref{eq:optimize_precoding}. Specifically , the Lagrangian function is given by 
\begin{equation}
   \mathfrak{L} = \sum_{k=1}^K \left(\vect{w}_k^\H \left(\vect{D}^\H \vect{D} + \mu \vect{I}_M \right) \vect{w}_k - \vect{d}_k^\T \vect{w}_k - (\vect{d}_k^\T \vect{w}_k)^\H \right) - \mu P,
\end{equation}where $\mu$ is the Lagrange multiplier associated with the power constraint. For a given $\mu$, the precoding optimization problem can be split into $K$ separate sub-problems. In particular, the problem for optimizing the $k$th precoding vector is given by 
\begin{equation}
\label{eq:minimize_lagrangian}
    \minimize{\vect{w}_k \in \mathcal{P}^M}\,\, \vect{w}_k^\H \left(\vect{D}^\H \vect{D} + \mu \vect{I}_M \right) \vect{w}_k - \vect{d}_k^\T \vect{w}_k - (\vect{d}_k^\T \vect{w}_k)^\H.
\end{equation}
Ignoring the fronthaul quantization, problem \eqref{eq:minimize_lagrangian} is a convex problem for which the optimal solution is readily obtained as 
\begin{equation}
\label{eq:precoding_without_quan}
   \hat{\vect{w}}_k  = \left(\vect{D}^\H \vect{D} + \mu \vect{I}_M \right)^{-1} \vect{d}^*. 
\end{equation}
If the precoding vector obtained in \eqref{eq:precoding_without_quan} is transmitted via the limited-capacity fronthaul to the BS, the real and imaginary parts of each of its entries are mapped to their nearest point in $\mathcal{L}$ as described in Section~\ref{sec:SysMod}. This uncontrolled quantization effect increases the inter-user interference and reduces the beamforming gain. We therefore propose an alternative approach that finds the optimal solution to \eqref{eq:minimize_lagrangian} while accounting for the fronthaul limitation. Specifically, defining $\vect{V} = \vect{D}^\H \vect{D} + \mu \vect{I}_M$ and decomposing $\vect{V}$ using Cholesky factorization as $\vect{V} = \vect{R}^\H \vect{R}$, we re-write \eqref{eq:minimize_lagrangian} as 
\begin{equation}
\label{eq:SESD_precoding}
   \minimize{\vect{w}_k \in \mathcal{P}^M}\,\, \|\vect{c}_k - \vect{R}\vect{w}_k \|^2 - \vect{c}_k^\H \vect{c}_k,
\end{equation}where $\vect{R} \in \mathbb{C}^{M\times M}$ is an upper-triangular matrix and $\vect{c} = (\vect{d}_k^\T \vect{R}^{-1})^\H$. Since $\vect{R}$ has a triangular structure, we can solve \eqref{eq:SESD_precoding} using the classical SESD algorithm \cite{agrell2002closest}. The optimal value of $\mu$
which provides a solution that satisfies the power constraint near equality can be found via a bisection search. 

\subsection{Optimizing RIS Configuration}
We now proceed to optimize the RIS configuration having the receiver gains and precoding vectors fixed. The problem is formulated as 
\begin{align}
\label{eq:optimize_RIS}
 \minimize{\bl{\theta} \in \mathcal{F}^N}\,\, &\bl{\theta}^\H \left(\sum_{k=1}^K \vect{F}_k^* \vect{W}^* \vect{W}^\T \vect{F}_k^\T  \right)\bl{\theta}  \nonumber \\
 - &\bl{\theta}^\H \left( \sum_{k=1}^K \vect{F}_k^* \vect{w}_k^*\right) - \left(\sum_{k=1}^K \vect{w}_k^\T \vect{F}_k^\T \right)\bl{\theta},
\end{align}where $\vect{F}_k = \diag(\vect{g}_k)\vect{H}$. A simple approach to solve problem~\eqref{eq:optimize_RIS} is to first drop the discrete reflection coefficient constraint, solve the problem for the continuous RIS reflection coefficients, and then map each obtained reflection coefficient to its nearest point in $\mathcal{F}$ \cite{Wu2019b}. To solve problem~\eqref{eq:optimize_RIS} for continuous reflection coefficients, alternating optimization can be utilized to alternately optimize one of the reflection coefficients having others fixed. Using this approach, the optimal reflection coefficient of the $n$th RIS element in each iteration is obtained as 
\begin{equation}
    \hat{\theta}_ n = - e^{-j \arg \left(\sum_{m \neq n} \theta_m^* [\vect{A}]_{m,n} - [\vect{a}]^*_n \right)}, 
\end{equation} 
where $\vect{A} = \sum_{k=1}^K \vect{F}_k^* \vect{W}^* \vect{W}^\T \vect{F}_k^\T,~~\vect{a} = \sum_{k=1}^K \vect{F}_k^* \vect{w}_k^*$.
$[\vect{A}]_{m,n}$ indicates the entry in the $m$th row and $n$th column of $\vect{A}$, and $[\vect{a}]_n$ represents the $n$th entry of $\vect{a}$.
The reflection coefficients are alternately optimized until a satisfactory convergence is achieved, i.e., after the difference between the objective function value of \eqref{eq:optimize_RIS} in two successive iterations becomes negligible. Assume that the obtained RIS reflection coefficient of the $n$th element after convergence is $\Tilde{\theta}_n$. The corresponding discrete reflection coefficient can then be obtained as 
\begin{equation}
   \overline{\theta}_n = \argmin{\theta \in \mathcal{F}}\,\, |\tilde{\theta}_n - \theta|. 
\end{equation}This method of finding the discrete RIS reflection coefficients has a low complexity; however, the solution is sub-optimal because the reflection coefficients are selected independently and the quantization errors pile up. To simultaneously optimize all RIS reflection coefficients, we propose to use the SESD algorithm and directly find the optimal discrete RIS configuration vector instead of first finding the continuous configuration and then quantizing it. 
To this end, the objective function of  problem~\eqref{eq:optimize_RIS} is re-written as 
\begin{equation}
  \bl{\theta}^\H \vect{A} \vect{\theta} - \bl{\theta}^\H \vect{a} - \vect{a}^\H \bl{\theta} =  \|\vect{b} - \vect{B}\bl{\theta}\|^2 - \vect{b}^\H \vect{b},
\end{equation}
where $\vect{B} \in \mathbb{C}^{N \times N}$ is an upper-triangular matrix obtained from Cholesky factorization of $\vect{A}$ as $\vect{A} = \vect{B}^\H \vect{B}$ and $\vect{b} = \left(\vect{a}^\H \vect{B}^{-1} \right)^\H$. Disregarding the term $\vect{b}^\H \vect{b}$ which is independent of $\bl{\theta}$, the problem for optimizing the RIS configuration is re-formulated as 
\begin{equation}
\label{eq:optimize_RIS_SD}
  \minimize{\bl{\theta} \in \mathcal{F}^N}\,\, \|\vect{b} - \vect{B}\bl{\theta}\|^2.
\end{equation}
 In practice, we normally have $K \leq M < N$, i.e., the number of UEs is less than the number of BS antennas and the number of BS antennas is less than the number of RIS elements. In this case, we have $\mathrm{rank}\left(\vect{F}_k^* \vect{W}^* \vect{W}^\T \vect{F}_k^\T \right)  \leq K,~k = 1,\ldots,K$. Therefore, $\mathrm{rank}(\vect{A}) \leq K^2$.
 If $K^2 < N$, then matrix $\vect{A}$ turns out to be rank-deficient which means that the Cholesky factor $\vect{B}$ is not full-rank and zero-valued entries appear on its diagonal. 
 Hence, the standard SESD algorithm is not applicable to solve \eqref{eq:optimize_RIS_SD}. Since $\|\bl{\theta}\|^2 = \bl{\theta}^\H \bl{\theta} = N $ is a constant, we can add the term $\alpha \bl{\theta}^\H \bl{\theta}$ to the objective function of \eqref{eq:optimize_RIS} without affecting its optimal solution. In particular, the objective function of \eqref{eq:optimize_RIS} can be modified as 
 \begin{equation}
  \bl{\theta}^\H \left(\vect{A} + \alpha \vect{I}_N \right) \vect{\theta} - \bl{\theta}^\H \vect{a} - \vect{a}^\H \bl{\theta} =  \|\tilde{\vect{b}} - \tilde{\vect{B}}\bl{\theta}\|^2 - \tilde{\vect{b}}^\H \tilde{\vect{b}},
\end{equation}
in which $\Tilde{\vect{B}}$ is upper-triangular such that $\vect{A} + \alpha \vect{I} = \tilde{\vect{B}}^\H \tilde{\vect{B}} $ and $\tilde{\vect{b}} = \left(\vect{a}^\H \tilde{\vect{B}}^{-1} \right)^\H$. As $\vect{A} + \alpha \vect{I}_N$ is full-rank, all the diagonal entries of $\Tilde{\vect{B}}$ are non-zero. Therefore, the SESD algorithm can be used to solve the  problem 
\begin{equation}
\label{eq:SESD_RIS}
 \minimize{\bl{\theta} \in \mathcal{F}^N}\,\, \|\Tilde{\vect{b}} - \Tilde{\vect{B}}\bl{\theta}\|^2.  
\end{equation}
The parameter $\alpha$ can be tuned to minimize complexity, but it is hard to obtain the optimal value  \cite{cui2004efficient}. In our simulations, we have used $\alpha = 1$. 

\subsection{Optimizing Receiver Gains}
To optimize the receiver gain $\beta_k$, we need to solve the problem
\begin{align}
\label{eq:optimize_beta}
   \minimize{\beta_k \in \mathbb{C}}\,\,e_k &=|\beta_k|^2 \left ( \sum_{i=1}^K \left| \bl{\theta}^\T \vect{F}_k\vect{w}_i\right|^2 + N_0 \right) \nonumber \\ 
 &- 2\Re \left ( \beta_k \bl{\theta}^\T \vect{F}_k \vect{w}_k \right ) + 1.  
\end{align}
We can see that the objective function is quadratic with respect to $\beta_k$. By taking the derivative of the objective function and equating it to zero, the solution to \eqref{eq:optimize_beta} is obtained as 
\begin{equation}
   \label{eq:optimal_beta}
       \overline{\beta}_k = \frac{\left(\boldsymbol{\theta}^{\Ttran}\vect{F}_k\vect{w}_k\right)^*}{ \sum_{i=1}^K |\boldsymbol{\theta}^{\Ttran}\vect{F}_k \vect{w}_i|^2 + N_0}.
   \end{equation}
   
Using the methods described above, we alternately optimize precoding vectors, RIS configuration, and receiver gains until the difference in sum MSE in two consecutive iterations becomes less than a predefined threshold.
Algorithm~\ref{Alg:BCD} summarizes the procedure for solving the sum MSE minimization problem in \eqref{eq:main_problem} using a block coordinate descent approach.

\begin{algorithm}[t!]
\caption{The proposed method for solving \eqref{eq:main_problem}}
\label{Alg:BCD}
\begin{algorithmic}[1]
\STATEx {\textbf{Inputs}: channels $\vect{H}$, $\{\vect{g}_k\}$, noise power $N_0$, maximum transmit power $P$. }
\STATE{Initialize the RIS configuration $\overline{\boldsymbol{\theta}}^{(0)}$ and BS precoding $\overline{\vect{W}}^{(0)} = [\overline{\vect{w}}_1^{(0)},\ldots,\overline{\vect{w}}_K^{(0)}]$}
\STATE{Compute $\{\overline{\beta}_k^{(0)}\}$ from \eqref{eq:optimal_beta} using  
$\overline{\boldsymbol{\theta}}^{(0)}$ and $\overline{\vect{W}}^{(0)} $ }
\STATE{Given $\overline{\vect{W}}^{(0)}$, $\overline{\boldsymbol{\theta}}^{(0)}$, and $\overline{\boldsymbol{\beta}}^{0}$, compute the sum MSE as $e^{(0)}$}
\STATE{Set the convergence threshold $\epsilon>0$}
\STATE{Set $\delta \gets \epsilon + 1$, $l \gets 0$}
\WHILE{$\delta > \epsilon$}
\STATE{$l \gets l+1$}
\STATE{Find the optimal precoding vectors $\{\overline{\vect{w}}_k^{(l)}\}$ by solving \eqref{eq:SESD_precoding} using the SESD algorithm for a given $\mu$; update $\mu$ using the bisection method and optimize the precoding vectors; do this until $\left|\sum_{k=1}^K \overline{\vect{w}}_k^{(l)\H} \overline{\vect{w}}_k^{(l)} - P\right| \leq \varepsilon$}
\STATE{Find the optimal RIS configuration $\overline{\boldsymbol{\theta}}^{(l)}$ by solving \eqref{eq:SESD_RIS} using the SESD algorithm}
\STATE{Find the optimal receiver gains $\{\overline{\beta}_k^{(l)}\}$ from \eqref{eq:optimal_beta}}
\STATE{Compute $\{e_k^{(l)}\}$ by substituting $\{\overline{\vect{w}}_k^{(l)}\}$,  $\overline{\boldsymbol{\theta}}^{(l)}$, and $\{\overline{\beta}_k^{(l)}\}$ into the MSE expression \eqref{eq:MSE} }
\STATE{Compute $e^{(l)} = \sum_{k=1}^K  e_k^{(l)}$}
\STATE{Set $\delta \gets |e^{(l)} - e^{(l-1)}|$}
\ENDWHILE
\STATEx{ \textbf{Outputs:} $\overline{\vect{w}}_k^{(l)}$, $\overline{\beta}_k^{(l)}$ $k = 1,\ldots, K,$ and $\overline{\boldsymbol{\theta}}^{(l)}$} 
 \end{algorithmic}
\end{algorithm}

\section{Numerical Simulations}
In this section, we will evaluate the performance of the proposed algorithm via Monte Carlo simulations. We will first show the convergence behavior and then evaluate the end-user performance by comparing the sum rate of the proposed scheme with benchmarks. Specifically, the rate of $\mathrm{U}_k$ is calculated as $\log_2(1+\mathrm{SINR}_k)$, where $\mathrm{SINR}_k$ is the signal-to-interference-plus-noise ratio at $\mathrm{U}_k$, given by 
\begin{equation}
    \mathrm{SINR}_k (\vect{W},\boldsymbol{\theta}) = \frac{|\boldsymbol{\theta}^{\Ttran} \vect{F}_k\vect{w}_k|^2}{\sum_{i=1,i\neq k}^K |\boldsymbol{\theta}^{\Ttran}\vect{F}_k \vect{w}_i|^2 + N_0}.
\end{equation}
Three benchmarks are considered for our evaluation:
\begin{enumerate} 
    \item \textit{SESD-based precoding only}: The precoding design is based on the SESD algorithm, while the RIS configuration design follows the conventional nearest point mapping method.  
    \item \textit{SESD-based RIS only}: The RIS design is based on the SESD algorithm while the precoding is obtained by quantizing the continuous precoding vectors  using the quantizer function $\mathcal{Q}(\cdot)$.
    \item  \textit{No SESD}: The RIS design is based on the nearest point mapping and the discrete precoding is obtained by quantizing the optimal continuous precoding  using the quantizer function $\mathcal{Q}(\cdot)$.
\end{enumerate}
We initialize the RIS configuration randomly and use the infinite-resolution regularized zero-forcing precoding for initializing the precoding vectors \cite{Joham} as $\overline{\vect{W}}^{(0)} = \Tilde{\vect{H}}^\H \left( \Tilde{\vect{H}} \Tilde{\vect{H}}^\H + \frac{KN_0}{P}\vect{I}_K\right)^{-1}$
where $\Tilde{\vect{H}} = \left(\vect{I}_K \otimes \overline{\bl{\theta}}^{(0)\Ttran}\right) \vect{GH}$ and $\overline{\bl{\theta}}^{(0)}$ is the initial RIS configuration vector. 
\vspace{-5mm}
\subsection{Simulation Setup}
We consider a system with $M = 4$ antennas at the BS in the form of a uniform linear array and $K = 3$ UEs. The RIS has $N = 64$ elements which are arranged in a square uniform planar array structure with $N_{\mathrm{H}} = 8$ elements in each horizontal dimension and $N_{\mathrm{V}} = 8$ elements in each vertical dimension. The number of fronthaul quantization levels is assumed to be $L = 4$ and the phase shift resolution of RIS elements is set to be $b = 1$. All channels are modeled by Rician fading. For example, the channel between the BS and RIS is modeled as $\vect{H} = \sqrt{\rho}\left(\sqrt{\frac{\kappa}{\kappa + 1}} \vect{H}_{\mathrm{LOS}} + \sqrt{\frac{1}{\kappa + 1}} \vect{H}_{\mathrm{NLOS}} \right),$
where $\kappa$ denotes the Rician factor set as $\kappa = 3$. The subscripts $\mathrm{LOS}$ and $\mathrm{NLOS}$ represent the line-of-sight (LOS) and non-LOS (NLOS) components of the channels. In particular,  $\vect{H}_{\mathrm{LOS}} = \vect{r}_{\mathrm{BS}}(\Omega) \vect{r}_{\mathrm{RIS}}^{\Ttran}(\varphi_{\mathrm{AoA}},\phi_{\mathrm{AoA}})$, 
where $\vect{r}_{\mathrm{BS}}(\cdot)$ and $\vect{r}_{\mathrm{RIS}}(\cdot)$ denote the array response vectors with $\Omega$, $\varphi_{\mathrm{AoA}}$, and $\phi_{\mathrm{AoA}}$ representing the angle of departure (AoD) from the BS, azimuth angle of arrival (AoA) to the RIS, and elevation AoA to the RIS, respectively, set as $\Omega = \frac{\pi}{6}$, $\varphi_{\mathrm{AoA}} = - \frac{\pi}{3}$, and $\phi_{\mathrm{AoA}} = \frac{\pi}{6}$. The spacing between the RIS elements in both horizontal and vertical dimensions is set as $\Delta_{\mathrm{H}} = \Delta_{\mathrm{V}} = \frac{\lambda}{4}$ and the spacing between BS antennas is set as $\frac{\lambda}{2}$, where $\lambda$ is the wavelength.
For $\vect{H}_{\mathrm{NLOS}}$, we consider correlated Rayleigh fading and use the local scattering spatial correlation model with Gaussian distribution \cite{Demir2022channel}.
In particular, let us express the NLOS channel matrix as $\vect{H}_{\mathrm{NLOS}} = [\vect{h}_1,\ldots,\vect{h}_M]$ where $\vect{h}_m \in \mathbb{C}^{N \times 1}$ represents the NLOS channel between the $m$th BS antenna and RIS. Using the correlated Rayleigh fading, this channel is modeled as $\vect{h}_m \sim \CN (\vect{0},\vect{C})$ in which $\vect{C} \in \mathbb{C}^{N \times N}$ is the spatial correlation matrix whose $(n,n^\prime)$th entry is given by 
\begin{equation}
\begin{aligned}
    &[\vect{C}]_{n,n^\prime} = \\ &\int \int e^{j\frac{2\pi}{\lambda}\left(\Delta_{\mathrm{H}} (n_{\mathrm{H}}-n^\prime_{\mathrm{H}})\sin(\varphi_{\mathrm{AoA}}+\eta_{\mathrm{az}})\cos(\phi_{\mathrm{AoA}} + \eta_{\mathrm{el}} ) \right) } \times \\
    & e^{j\frac{2\pi}{\lambda}\left(\Delta_{\mathrm{V}}(n_{\mathrm{V}} -n^\prime_{\mathrm{V}}) \sin(\phi_{\mathrm{AoA}} + \eta_{\mathrm{el}})\right)} f(\eta_{\mathrm{az}}) f(\eta_{\mathrm{el}}) d\eta_{\mathrm{az}} d\eta_{\mathrm{el}} 
    \end{aligned}
\end{equation}where $f(\cdot)$ represents the probability density function, deviation from nominal angles are Gaussian distributed, i.e.,   $\eta_{\mathrm{az}}, \eta_{\mathrm{el}} \sim \mathcal{N}\left(0,(\frac{\pi}{12})^2\right)$, $n_{\mathrm{H}} (n^\prime_{\mathrm{H}}) = \mod \left(n (n^\prime)-1,N_{\mathrm{H}}\right) +1$,  and $n_{\mathrm{V}}(n^\prime_{\mathrm{V}}) = \left\lceil \frac{n(n^\prime)}{N_{\mathrm{H}}} \right\rceil$.
Furthermore, $\rho$ denotes the path-loss at the carrier frequency of $3$\,GHz and is modeled as \cite{Emil2020RISvsDF}
\begin{equation}
     \rho = -37.5 - 22 \log_{10} \big(d/1~ \mathrm{m}\big) ~~[\mathrm{dB}],
\end{equation} with $d$ being the distance between the BS and the RIS, set as $d=20\,$m.
The channels between the RIS and the UEs are modeled similarly. The UEs are uniformly distributed around the RIS with their distance to the RIS given by $\mathcal{U}[20\,\mathrm{m},40\,\mathrm{m}]$, and their azimuth and elevation AoD distributed as $\mathcal{U}[0,\frac{\pi}{3}]$ and $\mathcal{U}[-\frac{\pi}{12},0]$, respectively. Finally, the noise power is $N_0 = -100\,$dBm. 
\begin{figure}[t]
    \centering
    \includegraphics[width = 0.8\columnwidth]{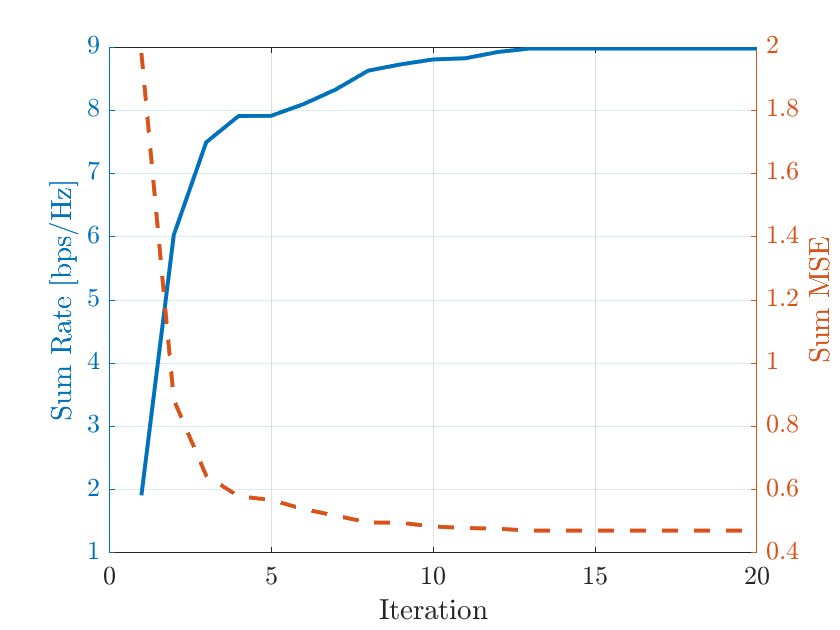}
    \caption{Convergence behavior of Algorithm~\ref{Alg:BCD}.}
    \label{fig:convergence}
\end{figure}
\begin{figure}[t]
    \centering
    \includegraphics[width = 0.8\columnwidth]{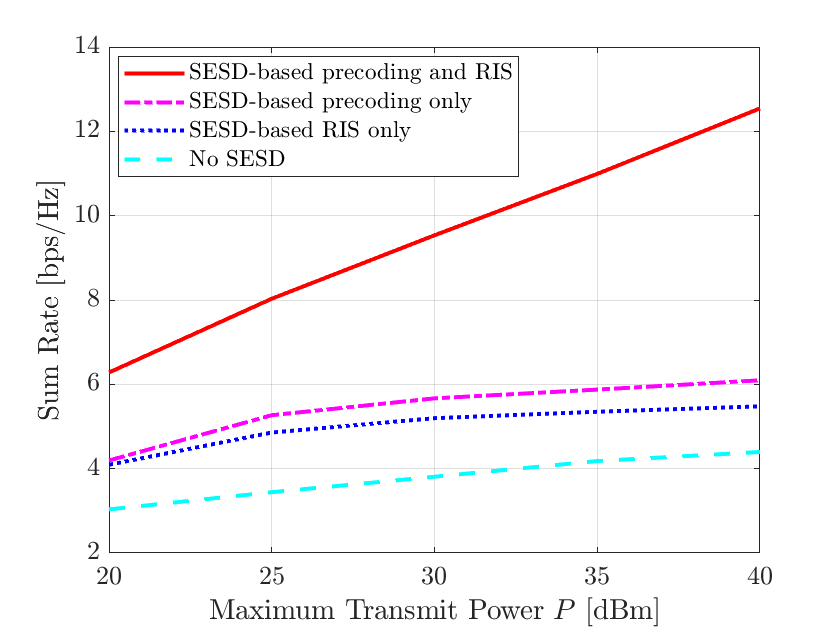}
    \caption{Sum rate versus maximum transmit power in an RIS-assisted MU-MIMO system with $M = 4$ BS antennas, $N = 64$ RIS elements, and $K = 3$ UEs. $L = 4$ quantization levels are used for precoding and $b = 1$ bit is used for RIS configuration.  }
    \label{fig:sumrate}
\end{figure}
\vspace{-3mm}
\subsection{Simulation Results}
Fig.~\ref{fig:convergence} shows the convergence behavior of the proposed Algorithm~\ref{Alg:BCD} in terms of both sum MSE and sum rate. The maximum transmit power of the BS is set as $P = 30\,$dBm. The proposed algorithm reaches a stationary point after about $13$ iterations. The algorithm guarantees convergence to a stationary point since we use block coordinate descent and the  SESD-based methods find the optimal precoding and RIS configurations in each iteration \cite{Tseng2001}. 

Fig.~\ref{fig:sumrate} plots the sum rate, as a function of the maximum transmit power, obtained by the proposed scheme and the three benchmarks discussed above. It can be seen that the proposed method, which designs both precoding and RIS configuration via the SESD algorithm, vastly outperforms all the benchmarks since it finds the optimal discrete precoding and RIS reflection coefficients. The gap between the proposed design and the benchmark schemes increases with increasing the transmit power because the uncontrolled quantization errors in benchmarks limit their interference suppression capability.   

\section{Conclusions}
The low bit resolution of RIS and the limited capacity of fronthaul link are two inherent properties of RIS-aided communication systems. Ignoring these two practical limitations when designing RIS configuration and precoding may result in severe performance degradation. 
In this paper, we investigated an RIS-assisted MU-MIMO downlink communication scenario and developed a novel framework for sum MSE minimization in which the discrete precoding vectors and RIS configuration have been optimized using the SESD algorithm. The proposed designs based on the SESD algorithm ensure that the optimal solution is selected from the discrete set which is in contrast to conventional sub-optimal methods where continuous solutions are first found and then quantized to the nearest discrete point. The interference mitigation offered by the proposed optimal designs improves the system performance over sub-optimal designs, as validated by numerical simulations.

\bibliographystyle{IEEEtran}
\bibliography{refs} 

\end{document}